# Saturation of the Raman amplification by self-phase modulation in silicon nanowaveguides

Felix Kroeger[1], Aleksandr Ryasnyanskiy[1], Alexandre Baron[1], Nicolas Dubreuil[1], Philippe Delaye[1], Robert Frey[1], Gérald Roosen[1], David Peyrade[2]

[1]*Laboratoire Charles Fabry de l'Institut d'Optique, CNRS, Univ Paris-Sud, Campus Polytechnique RD128, 91127 Palaiseau cedex, France*
[2]*Laboratoire des Technologies de la Microélectronique, CNRS, 17, rue des Martyrs, 38054 Grenoble Cedex 9, France*

**ABSTRACT :**

We experimentally show that the self-phase modulation of picosecond pump pulses, induced by both the optical Kerr effect and free-carrier refraction, has a detrimental effect on the maximum on-off Raman gain achievable in silicon on insulator nanowaveguides, causing it to saturate. A simple calculation of the Raman gain coefficient from the measured broadened output pump spectra perfectly matches the saturated behavior of the amplified Raman signal observed experimentally at different input pump powers.

Silicon-on-insulator (SOI) photonics has attracted a great deal of attention due to its potential solutions for on-chip optical data processing. The high refractive index combined with the large third-order nonlinearity exhibited by silicon[1], enable tight optical confinement in sub-micron waveguides and efficient nonlinear functionality, making SOI a promising platform for ultra-compact devices with low-command powers[2,3].

Stimulated Raman scattering (SRS) has been extensively studied for its applications to optical amplification, overcoming the indirect bandgap of silicon. Since SRS was first observed in SOI waveguides[4], Raman lasers operating in the pulsed[5] and continuous-wave[6] (CW) regimes have been demonstrated. As it turns out, the two-photon absorption (TPA) and the free-carrier absorption (FCA) induced by TPA constitute the two main limitations in achieving a high net Raman gain[2,3]. Because the phonon lifetime is close to 3 ps in crystalline silicon, pulses of several ps can be amplified efficiently using SRS at low repetition rates, corresponding to a regime in which sensitivity to FCA is strongly reduced.[7-10] However, depending on the pulse energy and modal confinement, both the optical Kerr effect and the free carriers can induce significant self-phase modulation (SPM) that leads to a considerable spectral broadening of the pump pulse[11]. When the broadening is of the order of magnitude of the Raman-gain bandwidth (105 GHz), a reduction of the gain coefficient is expected, which has not been reported in previous studies of the Raman amplification of ps pulses in SOI waveguides.[7-10]

In this letter, we report on the saturation of the Raman amlification due to the nonlinear spectral broadening experienced by ps pulses propagating in a L=11 mm long SOI nanowaveguide. The on-off Raman gain is measured in terms of the pump intensity and shows a clear saturation that is attributed to the spectral broadening of the pump pulse. Beyond the maximum 28 dB on-off Raman gain that is observed, we give a detailed experimental investigation of the interaction between SRS and nonlinear spectral broadening of the pump pulse in silicon waveguides. Hence, we provide a fundamental insight on the nonlinear interaction of short pulses in order to project and optimize the performance of future silicon-based data processing devices. Despite the tremendous amount of studies conducted on silicon Raman amplifiers and lasers, there is a large spread in the reported values of the Raman gain coefficient $\gamma_R$, which is quite remarkable[4,8-10]. Our results enable a measurement of $\gamma_R$ in relation to the TPA coefficient $\beta_{TPA}$. In addition, the relation between the Raman gain spectrum and the pump and signal spectral shapes is rarely taken into account, even though the broadening of ps pump pulses through SPM has been used to make a broadband Raman amplifier in silicon[12]. However, no comment is ever given on the expected saturation of the Raman amplification for a given signal wavelength.

Our experiments were carried out in ridge-type SOI nanowaveguides with a $A_{eff}$ = 500×340 nm² cross-section. Guiding occurs along the crystallographic direction [110]. They were fabricated on SOI wafers with 2-µm-thick oxide layers.[13] The Raman pump pulses are $\tau$ =15 ps in duration, with a repetition rate of $F$ = 80 MHz and are delivered by an optical parametric oscillator (OPO) synchronously pumped by a mode-locked Ti:Sapphire laser.[14] In order to investigate the Raman gain, we produce a pulsed signal synchronized with the pump pulse using a tunable CW laser diode source externally modulated by a LiNbO$_3$ intensity modulator. The modulation signal is provided by an amplified InGaAs photodetector with a 12 GHz bandwidth that detects a portion of the pump beam. We are able to produce 150 ps signal pulses, tunable around 1550 nm. The temporal and spatial overlapping between the pump and signal pulses in the waveguide, are achieved using a delay line and a dichroic beam splitter respectively. The pulses are launched into the nanowaveguide using a microscope objective with a numerical aperture of 0.85. A similar objective is used to collimate the transmitted beam at the output of the waveguide. The beam is then injected into a single-mode fiber, linked to an optical spectrum analyzer (OSA) with a 0.01 nm spectral resolution. Experimentally, the pump (resp. signal) pulse is injected in the TE (resp. TM) mode, one of the configurations for which the Raman susceptibility is maximum.[4]

Fig. 1 shows the spectral behavior of both the outgoing pump (a) and signal (b) pulses for increasing input pump average powers, measured before the microscope objective. The input pump wavelength is set to 1441 nm whereas the signal wavelength and time delay are adjusted so as to maximize the Raman amplification. Fig. 1(a) shows the pump spectrum broadened due to SPM induced by Kerr effect and free-carrier refraction (FCR). The instantaneous Kerr effect symmetrically broadens the spectrum, while the time-dependent buildup of the carrier density creates a nonlinear phase contribution, which produces a blue shift in the spectrum.[11] The spectral broadening induced by the Kerr and the FCR[3] is expected to be on the order of 1-2 nm for input intensities on the order of a few GW/cm², typically used in Raman amplification experiments. This is consistent with our observations. Dispersion effects can be neglected. Indeed, with the typical dispersion of ridge waveguides[3] ($\beta_2 \approx$ -1ps$^2$/m) we have a dispersion length $L_D = \tau_2^{\ 2}/|\beta_2|$=100 m much larger than the waveguide length $L$. The signal spectrum, shown in Fig. 1(b) at low input pump power, consists in a narrow peak for which the linewidth is limited by the spectral resolution of the OSA. As the pump power increases, one expects that only the overlapping part of the 150 ps signal pulse with the 15 ps pump pulse will be amplified. Consequently, the signal spectrum measured at the output of the waveguide should be broader. This feature is clearly seen in Fig. 1(b) where the output signal spectra present a broad component, close to the narrow peak, which is subsequently amplified and blue shifted as the pump power increases. For an input pump power of 120 mW and without any signal, the blue shifted amplified spontaneous Raman spectrum showed a maximum at 1557.3 nm. In presence of the input signal and for the same pump power, the Raman amplification was maximized by adjusting the input signal wavelength to 1558.2 nm for which the blue shifted spectrum of the Raman amplified signal is also centered at 1557.3 nm (see the top spectrum in Fig. 1(b)).

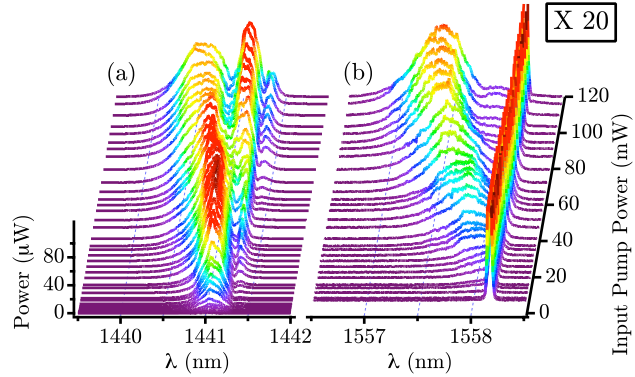

*FIG. 1 Output spectra of the pump pulses (a) and amplified signal pulses (b) measured at input average pump powers varying from 10 mW to 120 mW (the signal spectra should be divided by 20). The spectral resolution is 0.01 nm.*

In order to discuss the influence of the SPM-broadened pump pulse on SRS, we simply consider a CW signal at frequency $\omega_S$ amplified by a pump pulse. The coupled equations for the spatial evolution of the pump and signal intensities, under the undepleted pump approximation for SRS, are:

$$\begin{aligned}\frac{dI_P}{dz} &= -f_P^4 \beta_{TPA} I_P^2 - \alpha_P I_P, \\ \frac{dI_S}{dz} &= f_P^2 f_S^2 \left( \gamma_R(\omega_S) - \beta_{XTPA} \right) I_P I_S - \alpha_S I_S,\end{aligned} \quad (1)$$

where $\alpha_{P,S}$ accounts for linear losses for the pump and the signal, $\gamma_R(\omega_S)$ is the value of the Raman coefficient at $\omega_S$ and $\beta_{XTPA}$ is the cross-TPA coefficient. In our experiment, typical input pump intensities are on the order of a few GW/cm$^2$, with 15-ps pulses and a repetition rate of 80MHz, so following the arguments provided by *L. Yin et al.*[11], FCA can reasonably be neglected. This allows a simple analytical solution for Eq. (1). We have introduced in Eq. (1) a local field factor correction $f_P$, and $f_S$, respectively for the pump and signal amplitude fields, in order to take into account the local field enhancement induced by the group velocity reduction of the guided modes [15,16]. The solution of Eq. (1) provides:

$$I_P(0)/I_P(L) = (1+X)e^{\alpha_P L}, \qquad (2)$$

$$\ln[I_S(L)] = \frac{f_S^2(\gamma_R(\omega_S) - \beta_{XTPA})}{f_P^2 \beta_{TPA}} \ln[1+X] + \ln[I_S(0)e^{-\alpha_S L}], \quad (3)$$

where the parameter $X = f_P^4 \beta_{TPA} L_{eff} I_P(0)$ is defined in terms of the effective length of the waveguide $L_{eff} = (1-e^{-\alpha_P L})/\alpha_P$. The solution in Eq. (3) shows that the on-off Raman gain, defined as $G_{on-off} = I_S(L)/I_S(0)e^{-\alpha_S L}$, depends linearly on the 1+$X$ parameter on a log-log scale. However, the Raman gain coefficient $\gamma_R(\omega_S)$ directly depends on the broadened pump spectrum. Consequently, as the pump intensity increases, we expect that the experimental curve will deviate from the linear regime and exhibit a saturation of the on-off Raman gain.

In order to emphasize this effect, we extract the experimental values of $X$ from the input $P_{Pin}$ and output $P_{Pout}$ average pump powers that are measured outside the nanowaveguide. The total output pump powers $P_{Pout}$ are calculated using the spectra from Fig. 1(a). They are related to the internal pump intensity $I_P(L)$ by $P_{Pout}/(A_{eff} F\tau) = \kappa_{out} I_P(L)$, with $\kappa_{out}$ the output coupling efficiency, which includes the whole efficiency from the waveguide output to the OSA. A similar expression exists for the input pump power measured before injection: $P_{Pin}/(A_{eff} F\tau) = I_P(0)/\kappa_{in}$, with $\kappa_{in}$ the input coupling efficiency. We plotted in Fig. (2) the ratio $P_{Pin}/P_{Pout}$ vs the average incident pump power $P_{Pin}$. The linear dependence confirms that TPA only governs the pump depletion. In particular the pump depletion caused by SRS and FCA is negligible. Expressing Eq. (2) as $P_{Pin}/P_{Pout} = a + bP_{Pin}$, the linear fit gives $a$=3846±53 and $b$=103.97±0.93 mW$^{-1}$. These two parameters enable us to extrapolate an experimental value for $X$ by means of the relation: $X = bP_{Pin}/a$.

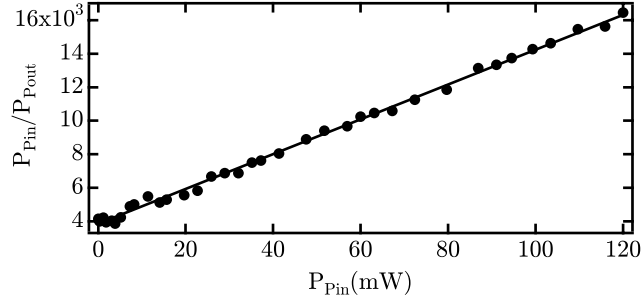

*FIG. 2 Plot of the ratio between the incident and output pump powers versus the incident pump power. The solid line corresponds to a linear fit of the experimental data.*

Concerning the on-off Raman gain behavior, we extract from the signal spectra in Fig. 1(b) the power of the amplified output signal $P_S(L) = \kappa_{out}(A_{eff}F\tau)I_S(L)$. The amplified components of the signal pulse can be easily filtered out from the optical spectra by integrating the power spectrum from 1556.5 nm up to 1558.1 nm, below the wavelength of the tunable laser. By doing so, we assume that the amplified signal spectrum is blue shifted. Indeed, no significant broadening of the narrow peak at 1558.2 nm is observable in the limit of the 0.01 nm OSA resolution. $P_{Sout}$ is then plotted on Fig. 3 as a function of $1+X$, using the previously determined values of $X$. For low input pump power, corresponding to $1+X$ <2dB, the amplified signal power linearly increases in a log-log scale as expected from Eq. (3). For higher input pump powers the experimental curve clearly deviates from the linear evolution and shows a saturation of the amplified signal. This saturation is not due to pump depletion caused by either TPA, which is included in Eq. (3), or SRS that is negligible. The solid line on Fig. 3 shows the linear fit of the experimental points below 2dB, from which we get a slope $s = 8.19 \pm 0.43$ and a y-intercept $P_0 = -62.63 \pm 0.66$ dBm. The latter gives the power of the output signal without pump that is temporally overlapping the pump pulse: $P_{Sout} = \kappa_{out}(A_{eff}F\tau)I_S(0)e^{-\alpha_S L}$. Using this value, we can now calculate the on-off Raman gain $G_{on-off}$ that is given on the right axis in Fig 3. A maximum value of 28 dB for the on-off Raman gain is achieved in this experiment, far below the ideal case of an un-broadened pump spectrum given by the solid line in Fig. 3.

The slope value $s$ provides the Raman gain coefficient $\gamma_R(\omega_S)$ assuming that the coefficients $f_S$, $f_P$, $\beta_{XTPA}$ and $\beta_{TPA}$ are known (see Eq. (3)). In the case of our nanowaveguide geometry, we can reasonably assume that $f_S \approx f_P$, because the variation between the pump and signal group velocities is expected to be less than 2%[2]. The value of $\beta_{TPA}$ in silicon around 1.5 μm is taken equal to 0.8 ± 0.12 cm/GW[1] and neglecting its dispersion one gets $\beta_{XTPA} = 2\beta_{TPA}$. The Raman gain coefficient is then estimated to $\gamma_R(\omega_S)$ = 8.1 ± 1.6 cm/GW. Lastly, we calculated the on-off Raman gain that can be expected from the measured spectral lineshapes of the pump. Assuming the signal to be CW at $\omega_S$, $\gamma_R(\omega_S)$ is given by the convolution product between the Raman gain spectrum $H_R(\Omega)$ and the spectral density $S_P(\omega)$ of the pump. The Raman gain spectrum is described by a Lorentzian lineshape with a full width at half maximum equal to 105 GHz[4]. For low input pump powers, for which no significant broadening is observable, the convolution product shows a reduction of the maximum Raman gain coefficient of 10% compared to the steady-state value $\gamma^0_R$. This means that the duration of the pump (15 ps) corresponds to a slightly transient regime. So, we set $\gamma^0_R$ to 8.9 cm/GW and calculate for each pump power the value of $\gamma_R(\omega_S)$ from the convolution product. The calculated values of $G_{on\text{-}off}$ are then plotted with open squares in Fig. 3. Despite its simplicity, the model fits the experimental data remarkably well and demonstrates that the nonlinear spectral broadening of the pump is responsible for the saturation observed in Fig. 3. This result is obtained by considering a fixed pump linewidth (measured at the output of the waveguide) without taking into account the spatial dependence of the pump broadening along the waveguide. A detailed numerical analysis of the Raman gain saturation that takes into account TPA, optical Kerr and free carrier effects, including FCA, is required and is presently under progress. Preliminary results show that most of the spectral broadening of the pump occurs on a typical propagation length of 2 mm. This length contributes to 1% only of the total Raman amplification.

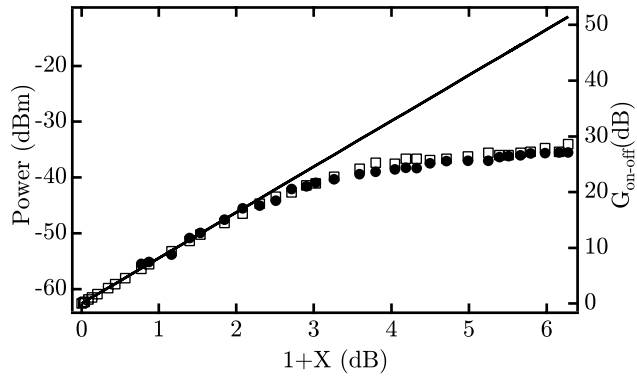

FIG. 3 Plot of the average power of the Raman amplified output signal in dBm with the corresponding on-off Raman gain (right axis) versus the 1+X parameter in dB (filled circles). The values directly calculated from the measured lineshape of the pump spectra are plotted with open squares. The solid line corresponds to the un-broadened pump case given by Eq. (3).

For information, we give hereafter a correspondence between the values taken by the parameter *X* and the injected pump intensity. Assuming a value for $\beta_{TPA}$, one needs to determine the linear attenuation and the group velocity of the waveguide. We performed a linear transmission spectroscopy of the nanowaveguide in the range 1540-1560 nm, using a tunable CW laser. The modulation depth and the period of the Fabry-Perot fringes, which originate from the reflection between the input and output facets of the waveguide, are respectively related to the linear attenuation and the group velocity. The estimated reflectivity of the facets is 30%, given by the Fresnel reflection coefficient for the interface air-silicon. The measured linear attenuation coefficient is then $\alpha_S = 1$ cm$^{-1}$. The group velocity is measured equal to c/5.1, which gives a local field factor $f_S = 1.2$, with a refractive index for the bulk silicon equal to 3.5. In the case of our experiment, and neglecting the variation of these values for the pump, we have $I_P(0) \approx 0.95X$ GW/cm$^2$. So, the experimental injected pump intensities reported on the bottom axis of Fig. 3 and Fig. 1b vary from 0.25 to 3 GW/cm$^2$ and the saturation occurs for injected pump intensities higher than 0.65 GW/cm$^2$. Notice that above 3 GW/cm$^2$, FCA is no longer negligible[11].

In conclusion, we have experimentally shown a saturation of the Raman gain in a SOI nanowaveguide in the ps regime, which is caused by nonlinear spectral broadening of the pump governed by both the optical Kerr and free carrier effects. With a simple model, using the lineshapes of the outgoing pump spectra, we reproduce the saturated behavior of the on-off Raman gain along with its maximum value (28 dB). Although this experiment was performed with a 15 ps pulse duration, corresponding to a slightly transient regime for SRS, this model enables the measurement of the steady-state Raman gain coefficient $\gamma^0_R = 8.9$ cm/GW for silicon, which is in accordance with the values found in the literature. Beyond the measurement method proposed, this work provides insight in the study of slow-mode Raman devices. Indeed, increasing the group index of both pump and signal beams, for the purpose of compactness and low-command power, will be detrimental to SRS because of the simultaneous enhancement of the spectral broadening of the pulses, in particular through the phase modulation effects[15]. Accurate understanding of the interaction between SRS and phase modulation is crucial in the study of Raman functionality.

This work was supported by the ANR. The authors thank Prof. Govind P. Agrawal for helpful discussions.